\documentclass[conference]{IEEEtran}
\IEEEoverridecommandlockouts

\usepackage{cite}
\usepackage{amsmath,amssymb,amsfonts}
\usepackage{algorithmic}
\usepackage{graphicx}
\usepackage{textcomp}
\usepackage{xcolor}
\usepackage{hyperref}
\def\BibTeX{{\rm B\kern-.05em{\sc i\kern-.025em b}\kern-.08em
    T\kern-.1667em\lower.7ex\hbox{E}\kern-.125emX}}
\begin{document}

\title{DeFT-Mamba: Universal Multichannel Sound Separation and Polyphonic Audio Classification
\thanks{This work was supported by the National Research Foundation of Korea (NRF) grant funded by the Ministry of Science and ICT of Korea government (MSIT) (No. RS-2024-00337945), the BK21 FOUR program through the NRF grant funded by the Ministry of Education of Korea government (MOE), and the Center for Applied Research in Artificial Intelligence (CARAI) funded by DAPA and ADD (UD230017TD)}
}

\author{\IEEEauthorblockN{Dongheon Lee}
\IEEEauthorblockA{\textit{Electrical Engineering} \\
\textit{Korea Advanced Institute of Science and Technology}\\
Daejeon, Republic of Korea \\
donghen0115@kaist.ac.kr}
\and
\IEEEauthorblockN{Jung-Woo Choi}
\IEEEauthorblockA{\textit{Electrical Engineering} \\
\textit{Korea Advanced Institute of Science and Technology}\\
Daejeon, Republic of Korea \\
jwoo@kaist.ac.kr}
}

\maketitle

\begin{abstract}
This paper presents a framework for universal sound separation and polyphonic audio classification, addressing the challenges of separating and classifying individual sound sources in a multichannel mixture. The proposed framework, DeFT-Mamba, utilizes the dense frequency-time attentive network (DeFTAN) combined with Mamba to extract sound objects, capturing the local time-frequency relations through gated convolution block and the global time-frequency relations through position-wise Hybrid Mamba. DeFT-Mamba surpasses existing separation and classification networks by a large margin, particularly in complex scenarios involving in-class polyphony. Additionally, a classification-based source counting method is introduced to identify the presence of multiple sources, outperforming conventional threshold-based approaches. Separation refinement tuning is also proposed to improve performance further. The proposed framework is trained and tested on a multichannel universal sound separation dataset developed in this work, designed to mimic realistic environments with moving sources and varying onsets and offsets of polyphonic events.
\end{abstract}

\begin{IEEEkeywords}
multichannel universal sound separation, polyphonic audio classification, mamba, self-attention
\end{IEEEkeywords}

\begin{figure*}[t]
    \vspace{-1em}
    \centering
    \includegraphics[width=0.95\textwidth]{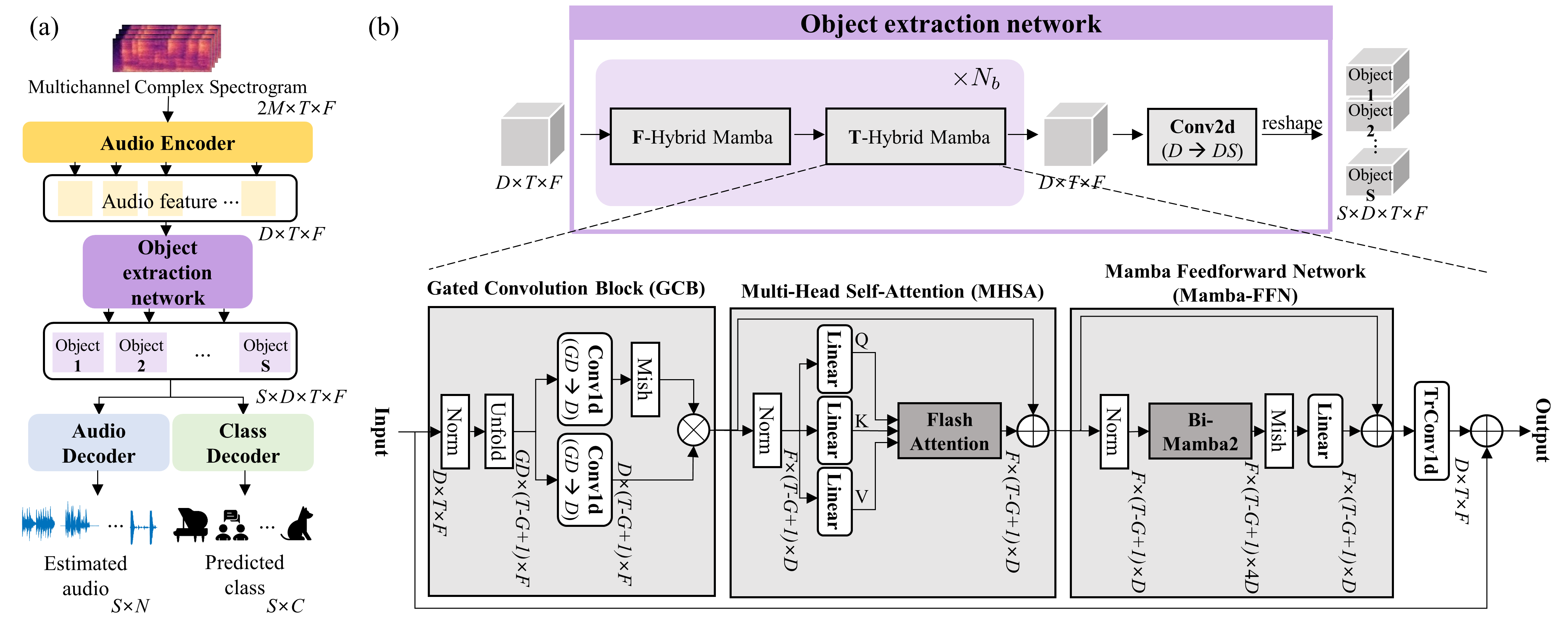}
    \vspace{-0.5em}
    \caption{(a) Proposed framework for universal sound separation and polyphonic audio classification, and (b) illustration of object extraction network. The Hybrid Mamba module consists of a gated convolution block (GCB), multi-head self-attention (MHSA), and Mamba feedforward network (Mamba-FFN)}
    \label{fig:framework}
    \vspace{-1em}
\end{figure*}

\section{Introduction}
The cocktail party problem involves identifying and separating individual sound sources in a noisy reverberant environment where multiple sound sources overlap \cite{haykin2005cocktail}. This problem has been a major focus in research areas such as computational auditory scene analysis (CASA) \cite{wang2006computational}. Recently, universal sound separation (USS), which focuses on separating all individual sound sources from a mixture of various signals, has gained significant research interest \cite{kavalerov2019universal, wisdom2021s, tzinis2020improving, liu2024audio, pons2024gass, luo2023music}. The diversity of source domains and the unknown number of source signals make USS more challenging than tasks that consider single-domain data such as speech separation \cite{kavalerov2019universal, wisdom2021s}.

Effectively addressing USS requires the development of comprehensive deep neural networks (DNNs) \cite{kavalerov2019universal, wisdom2021s, luo2023music} and compatible datasets \cite{wisdom2021s, fonseca2021fsd50k, petermann2023tackling, veluri2023real}. 
Recent state-of-the-art (SOTA) models for multichannel sound separation \cite{luo2023music, wang2023tf, quan2024spatialnet} employ convolutional or recurrent neural networks (CNNs or RNNs) to model and capture the spatial, spectral, and temporal characteristics of signals \cite{quan2024spatialnet, luo2023music, wang2023tf}. 
Transformers have also brought innovations in this field \cite{lee2023deft, quan2024spatialnet}. However, the computational complexity of transformer-based models increases quadratically with sequence length, and many alternatives have recently been proposed to address this issue \cite{chen23g_interspeech, zhao2024sicrn}. Mamba is one of such models designed to handle relations over long sequences \cite{gu2023mamba}, which has demonstrated its efficacy in speech separation tasks\cite{li2024spmamba, jiang2024dual, zhang2024mamba}. Leveraging such novel architecture in USS would also be beneficial. 

On the other hand, existing USS datasets \cite{wisdom2021s, petermann2023tackling, veluri2023real} have primarily focused on simulating mixtures with a variety of source classes and numbers, which only includes single-channel data of sources fixed in space. In datasets such as STARSS23 \cite{politis2023starss23}, moving sources with distinct onsets and offsets are considered for sound event localization and detection. However, ground-truth target source signals are not provided, making them unsuitable for the USS task. 

Besides separation, audio classification is also an essential task in CASA \cite{cakir2017convolutional, hershey2017cnn, gong2021ast, erol2024audio}. In particular, polyphonic audio classification is necessary for source counting and classification when multiple sources are overlapped \cite{cakir2017convolutional}. Especially, in-class polyphony, where sounds from the same class are overlapped, is a major cause of reduced classification performance \cite{mesaros2019sound}. To address the in-class polyphony issue, individual sounds need to be separated based on other cues before being classified.

This study proposes a unified model for the USS and polyphonic audio classification tasks. We introduce a framework that separates features for individual sound sources from a sound mixture and decodes them into audio and classes using sub-decoders. The contributions of this work can be summarized as follows: 
\begin{itemize}
    \item We propose Dense Frequency-Time Mamba (DeFT-Mamba), which captures local time-frequency relations using gated convolution blocks and global relations through self-attention, combined with position-wise Mamba to separate individual sound sources.
    \item We introduce a multichannel simulation dataset applicable to both USS and polyphonic audio classification tasks, featuring signals from moving sources with onset and offset variations.
    \item We incorporate a classification-based source-counting mechanism to improve source-counting accuracy and separation refinement tuning (SRT) to enhance separation performance using the estimated number of sources.
\end{itemize}

\section{Proposed method}
\subsection{Multi-task learning framework}
This work aims to separate individual sound source signals from a multichannel mixture recorded in noisy reverberant environments. The mixture can include up to $S$ foreground source signals, while the remaining signals are considered background noise that needs to be removed. The multichannel USS problem can be described as
\begin{equation}
Y_m(t, f) = \sum_{s=1}^{S}X_{m,s}(t, f)+V_m(t, f)\label{math_model},
\end{equation}
where $Y_m(t,f)$, $X_{m,s}(t,f)$, and $V_m(t,f)$ are multichannel spectrogram of a sound mixture, the reverberant sound of the $s$-th sound source, and noise captured by the $m$-th microphone, in the time frame $t=1,\cdots,T$ and frequency bin $f=1,\cdots,F$, respectively. 

The overview of the proposed multi-task learning framework for USS and polyphonic audio classification is illustrated in Fig. \ref{fig:framework}(a). The model takes a multichannel sound mixture recorded from an $M$-channel microphone array as input. The complex spectrogram of a multichannel sound mixture is mapped into a feature space by the audio encoder, which consists of a 2-D convolution and group layer normalization \cite{lee2023deftan}. The audio encoder increases the channel dimension from $2M$ to $D$. Then, the features are separated into $S$ tracks of object-dependent features by the object extraction network, followed by two parallel decoders that decode them into estimated source waveforms of length $N$ and predicted class probabilities for $C$ classes.
The audio decoder employs a 2D convolution that reduces the channel dimension from $D$ to $2$ to generate the real and imaginary (RI) components of the complex spectrogram, and the class decoder is a convolutional recurrent neural network (CRNN) \cite{cakir2017convolutional}, a common model for the audio classification task. The advantage of this framework is that the pairing of estimated waveforms and class probabilities is unnecessary because both are derived from the same features already separated by the object extraction network. 

\begin{table*}[t]
\caption{The statistics of duration (hours) and number of audio clips for each class of the proposed dataset}
\vspace{-0.5em}
\setlength{\tabcolsep}{3.4pt}
\scalebox{0.94}{
\begin{tabular}{c|ccccccccccccc}
\hline
Class                                                            & Female speech & Male speech & Clapping & Telephone & Laughter & Domestic sounds & Walk & Door & Music & Instrument & Water tap & Bell  & Knock \\ \hline
Number of clips & 14,219        & 14,320      & 632      & 710       & 1569     & 619             & 880  & 572  & 150   & 4,819              & 700       & 1,098 & 621   \\
Durations (h)                                                    & 50.0          & 50.6        & 1.84     & 1.61      & 1.86     & 2.60            & 2.32 & 0.73 & 9.81  & 7.70               & 2.67      & 2.79  & 0.73  \\ \hline
\end{tabular}}
\label{tab:data}
\vspace{-1em}
\end{table*}

\subsection{Dense Frequency-Time Mamba (DeFT-Mamba)}
The object extraction network plays a key role in extracting features for individual sound objects. We introduce a novel framework, DeFT-Mamba, which enhances the existing DeFTAN-II architecture by incorporating a gated convolution block (GCB) and Mamba2~\cite{dao2024transformers}. 
The overall architecture of DeFT-Mamba is illustrated in Fig. \ref{fig:framework}(b).

DeFT-Mamba consists of $N_b$ time(T)- and frequency(F)-Hybrid Mamba blocks. Each Hybrid Mamba block processes its input through GCB, an advanced version of the gated split dense block (GSDB) used in the previous work \cite{lee24g_interspeech} to capture local time-frequency relations. GCB shifts features by one time frame (T-Hybrid Mamba) or one frequency bin (F-Hybrid Mamba) using an unfold operation with a kernel size of $G$ and a stride of $1$. This is followed by two parallel 1-D convolution layers that reduce the channel dimension from $GD$ to $D$. GCB has a parallelizable structure and can shorten the long runtime required for GSDB without sacrificing performance.
A gating mechanism is then applied to selectively control the flow of information, with the Mish activation function applied to one branch to enhance gradient propagation \cite{misra2019mish}.

Next, the extracted features are processed by a multi-head self-attention (MHSA) mechanism. The self-attention module, implemented without positional embeddings, captures global relations independently of the sequential positions of feature embeddings. For efficacy, FlashAttention-2 (FA) \cite{dao2023flashattention} with linear projection is utilized in this work. MHSA is then followed by Mamba feedforward network (Mamba-FFN), designed to address position-wise global relationships. Mamba processes sequential input in a recurrent manner, which can be represented by the state-space equation~\cite{gu2023mamba}:
\begin{align}
h_k &= \Bar{A}h_{k-1} + \Bar{B}x_k \\
y_k &= \Bar{C}h_{k} + \Bar{D}x_k,
\end{align}
where $x_k, h_k, y_k$ denote the input, hidden state, output of $k$-th sequence (time for frequency), respectively, and $\Bar{A}, \Bar{B}, \Bar{C}, \Bar{D}$ represents the discretized state matrices. 
This recursive characteristic allows Mamba to model position-dependent global relations. Unlike position-wise feedforward networks using RNNs or gated recurrent units (GRUs)~\cite{chen2020dual, wang2021tstnn}, which do not support parallel processing, Mamba is parallelizable, thus facilitating an efficient position-wise feedforward network. 

After $N_b$ repetitions of F- and T-Hybrid Mamba blocks, a 2-D convolution is applied to aggregate and group the features for $S$ sound objects. This operation expands the channel dimension from $D$ to $DS$, which is then split into $S$ tracks (batches), each corresponding to an isolated sound object.

\begin{figure}[]
    \vspace{0em}
    \centering
    \includegraphics[width=0.95\columnwidth]{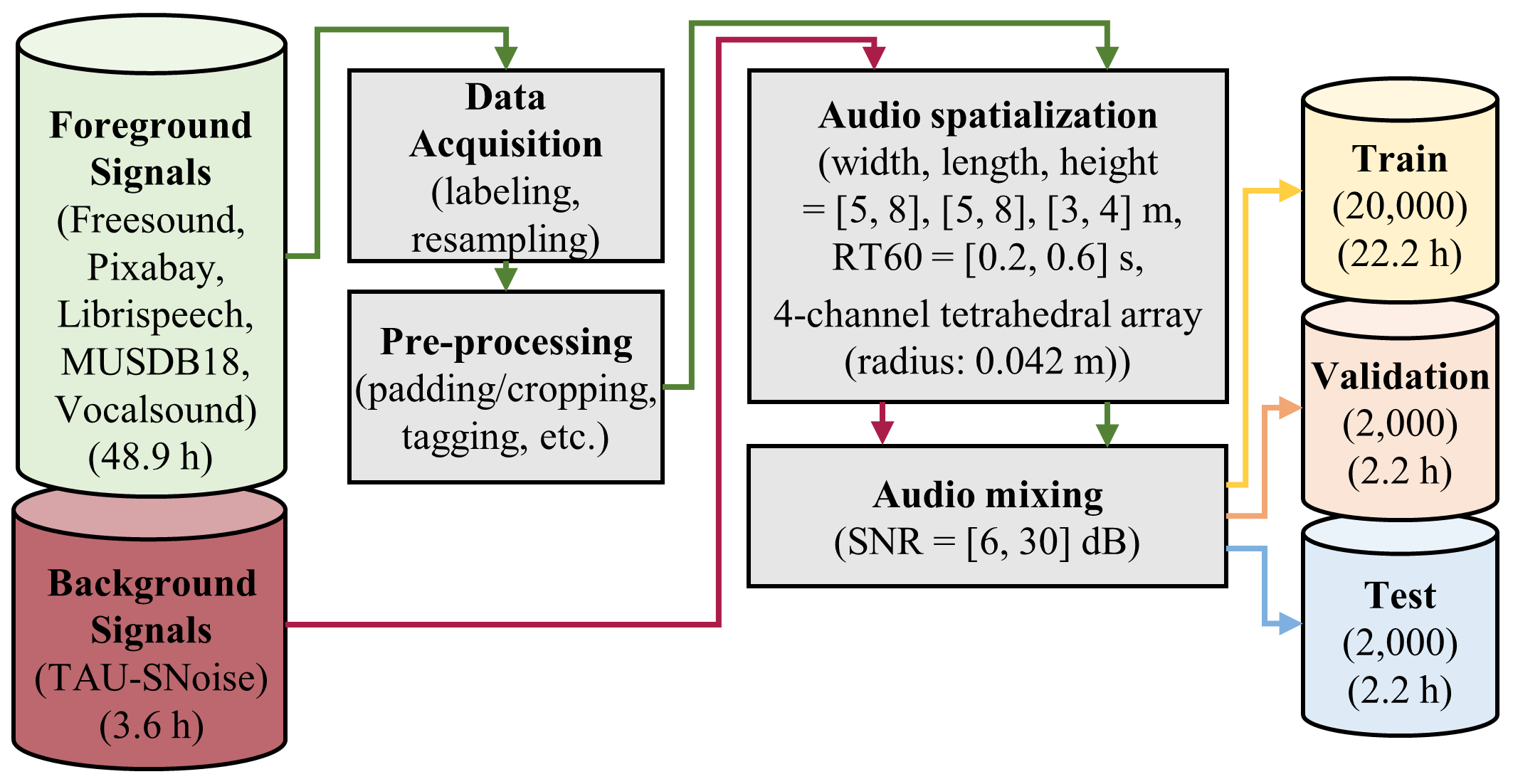}
    \vspace{-1em}
    \caption{Construction of simulation dataset.}
    \label{fig:dataset}
    \vspace{-1em}
\end{figure}

\section{Experiment}
\subsection{Dataset}
We constructed a new dataset for multichannel USS and polyphonic audio classification tasks. The proposed dataset is designed to reflect various conditions, including moving sources with temporal onsets and offsets. For foreground sound sources, signals from 13 audio classes were selected from open-source databases (Pixabay\footnote{\url{https://pixabay.com/sound-effects}} and FSD50K \cite{fonseca2021fsd50k}, Librispeech \cite{panayotov2015librispeech}, MUSDB18 \cite{rafii2017musdb18}, Vocalsound \cite{gong2022vocalsound}). These signals were resampled to 16 kHz 
and pre-processed by either padding zeros or cropping to 4 seconds. Each sound source has a 75\% probability of being a moving source, with speeds ranging from 0 to 3 m/s. The dataset features between 2 to 4 foreground sound sources, along with one background noise from the diffused TAU-SNoise dataset\footnote{\url{https://zenodo.org/records/6408611}} with a signal-to-noise ratio (SNR) ranging from 6 to 30 dB. The simulations were conducted using gpuRIR \cite{diaz2021gpurir}. Room dimensions were set to a width and length between 5 and 8 meters, and a height between 3 and 4 meters, with reverberation times ranging from 0.2 to 0.6 seconds. These parameters were sampled from uniform distributions. We simulated spatialized sound sources using a 4-channel tetrahedral microphone array with a radius of 4.2 cm. The procedure for dataset generation is illustrated in Fig. \ref{fig:dataset}, and details about class configuration and durations of audio clips are provided in Table \ref{tab:data}. This dataset poses a significant challenge for separation tasks due to the inclusion of moving sources, onset and offset conditions, overlapped in-class sources, and noisy reverberant environments. The dataset is available for download on the webpage\footnote{\url{https://zenodo.org/records/13749621}}.

\subsection{Training methods and parameter setup}
The training process consists of two stages. In the first stage, separation and classification were performed using a joint loss of source-aggregated signal-to-distortion ratio (SA-SDR) \cite{von2022sa} and cross-entropy loss. 
We chose SA-SDR loss over free universal sound separation (FUSS) loss \cite{wisdom2021s} due to its stability during training. The target classes for separation were 14 classes, including 13 classes for source signals and one silence class for mapping inactive sources. The source was classified as inactive when the 14th class had the highest probability, and as active for any other class. Permutation invariant training (PIT) \cite{yu2017permutation} was applied to separated source signals of $S$ tracks, with the permutation order shared between the separated signals and their predicted classes.

In the second stage, the estimated number of sound sources from the first stage was used for fine-tuning, denoted as separation refinement tuning (SRT). Since SA-SDR loss is not optimal for maximizing separation performance \cite{kim2023training}, a linear combination of 10\% scale-invariant signal-to-distortion ratio (SI-SDR) and 90\% signal-to-distortion ratio (SDR) was used as the loss function \cite{veluri2023real}. For SRT, the loss function is calculated only for the estimated number of sources after sorting the estimated sources according to the permutation determined from the first stage. 

For the short-time Fourier transform (STFT), a Hamming window with a 32 ms length and a hop size of 16 ms was employed. The number of DeFT-Mamba blocks was set to $N_b=6$, with a channel dimension $D=64$. The training was conducted using the ADAM optimizer with a learning rate of $1\times 10^{-4}$. The model was trained for 100 epochs in the first and 20 epochs in the second stages, respectively. Separation performance was evaluated using SI-SDR, SDR, and perceptual evaluation of audio quality (PEAQ) \cite{thiede2000peaq}\footnote{\url{https://github.com/stephencwelch/Perceptual-Coding-In-Python}}, while classification performance was assessed based on error rate (ER), F1 score, and source counting accuracy (SCA) \cite{kim2023training}. Additionally, to evaluate the efficiency of the model, we compared the parameter size (Param.), computational complexity (MAC/s), and training time per epoch (Runtime).

\section{Result}
We first conducted ablation studies to evaluate the effectiveness of each modification introduced in the proposed model, with particular attention to computational complexities. 
The results compared to the baseline speech enhancement model (DeFTAN-II (base)) are presented in Table \ref{tab:ablation}. Our first modification, the GCB replacing the split dense block (SDB), resulted in improved SI-SDR and SDR through the gating mechanism. Notably, this performance improvement was accomplished with a smaller parameter size, reduced computational complexity, and shorter runtime, through the parallelization of convolutions. 
The second modification of replacing the convolutional efficient attention (CEA) of DeFTAN-II with FA and linear projection showed similar performance but enhanced efficiency-related metrics (parameter size, computational complexity, runtime). We denote this model with FA and GCB as the second baseline: Model A. 

Next, Model A with the whole transformer blocks replaced by Mamba showed improved performance, similar to the results reported in other studies for single-channel speech separation \cite{li2024spmamba, jiang2024dual}. Performance was further enhanced when only MHSA was replaced by Mamba while keeping the feedforward network unchanged, which is in line with the result reported in \cite{zhang2024mamba}. 

The most striking improvement was brought by the proposed Hybrid Mamba structure (DeFT-Mamba) that uses both MHSA and Mamba-FFN simultaneously. To check if the Mamba-FFN just worked as another transformer, we compared the proposed model with the model using doubled transformer blocks (Model A (large)). Our model outperforms this in every performance metric by a large margin at much lower space and time complexities. 

\begin{table}[]
\caption{Ablation study results}
\vspace{-0.5em}
\setlength{\tabcolsep}{3.0pt}
\scalebox{0.96}{
\begin{tabular}{l|cc|ccc}
\hline
             & SI-SDR & SDR  & Param. & MAC/s   & Runtime  \\ \hline
unprocessed  & -5.48  & -5.54 & -      & -       & -        \\ \hline
DeFTAN-II (base)    & 4.11   & 6.95 & 4.1 M & 66.1 G  & 28.2 min \\
\ {\scriptsize replacing SDB with GCB}    & 4.28   & 7.09  & 3.9 M & 55.2 G  & 17.1 min \\
\ {\scriptsize + FA \& linear proj. (Model A)}  & 4.26   & 7.11 & 3.8 M & 54.8 G  & 14.0 min \\
\ {\scriptsize + transformer $\rightarrow$ Mamba}  & 4.38   & 7.19 & \textbf{3.0 M} & \textbf{41.2 G}  & \textbf{12.8 min} \\
\ {\scriptsize + MHSA $\rightarrow$ Mamba} & 4.51   & 7.21 & 3.8 M & 43.6 G  & 13.2 min \\
Model A (large)   & 4.52   & 7.05 & 7.3 M & 102.4 G & 28.5 min \\
\textbf{DeFT-Mamba} & \textbf{4.95}   & \textbf{7.27} & 4.2 M & 58.2 G  & 20.7 min \\ \hline
\end{tabular}}
\label{tab:ablation}
\vspace{-1em}
\end{table}

We compared the USS performance of DeFT-Mamba with SOTA models, including MC-BSRNN \cite{gu2024rezero}, TF-GridNet \cite{wang2023tf}, and SpatialNet \cite{quan2024spatialnet}, trained using the SA-SDR loss. The results in Table \ref{tab:comp_sep} indicate that DeFT-Mamba outperforms these existing models while using fewer parameters and maintaining comparable complexities. This demonstrates the effectiveness of the Hybrid Mamba in USS, particularly its ability to capture local time-frequency relations through GCB and global time-frequency relations via Mamba. Additionally, the model fine-tuned using SRT showed further improvements in SI-SDR, SDR, and PEAQ, suggesting that using the number of sound sources as additional information can further improve separation performance.

\begin{table}[]
\caption{Comparison with SOTA separation models}
\vspace{-0.5em}
\setlength{\tabcolsep}{3.4pt}
\scalebox{0.96}{
\begin{tabular}{c|ccc|ccc}
\hline
           & SI-SDR & SDR  & PEAQ & Param. & MAC/s   & Runtime  \\ \hline
MC-BSRNN \cite{gu2024rezero}   & 3.44   & 6.18 & -2.59   & 12.2 M & \textbf{15.3 G}  & \textbf{3.0 min}  \\
TF-GridNet \cite{wang2023tf} & 3.71   & 6.54 & -2.50   & 14.7 M & 462.2 G & 74.5 min \\
SpatialNet \cite{quan2024spatialnet} & 4.38   & 7.16 & -2.38   & 7.3 M  & 71.8 G  & 38.4 min \\
\textbf{DeFT-Mamba}  & \textbf{4.95}   & \textbf{7.27} & \textbf{-2.32}  & \textbf{4.2 M} & 58.2 G  & 20.7 min \\ \hline
\textbf{DeFT-Mamba+SRT} & \textbf{5.12} & \textbf{7.44} & \textbf{-2.27} & - & - & - \\ \hline
\end{tabular}}
\label{tab:comp_sep}
\vspace{-1em}
\end{table}

Furthermore, classification performance was compared with CRNN \cite{cakir2017convolutional}, Audio Spectrogram Transformer (AST) \cite{gong2021ast}, and AudioMamba (AuM) \cite{erol2024audio} (Table \ref{tab:comp_cls}). The proposed DeFT-Mamba demonstrates superior classification performance, indicating that separating individual sound sources before class prediction yields better results than directly classifying from a sound mixture in polyphonic audio classification.

\begin{table}[]
\centering
\caption{Comparison of classification performance}
\vspace{-0.5em}
\begin{tabular}{c|cc|ccc}
\hline
           & ER $\downarrow$ & F1 $\uparrow$  & Param. & MAC/s   & Runtime  \\ \hline
CRNN \cite{cakir2017convolutional}   & 74.3   & 20.9 & \textbf{1.8 M} & \textbf{2.3 G}  & \textbf{1.2 min}  \\
AST \cite{gong2021ast} & 70.6   & 25.8 & 85.3 M & 6.2 G & 2.8 min \\
AuM \cite{erol2024audio} & 68.4   & 27.0 & 90.9 M  & 17.2 G  & 2.5 min \\
\textbf{DeFT-Mamba}  & \textbf{31.8}   & \textbf{60.3} & 4.22 M & 58.2 G  & 20.7 min \\ \hline
\end{tabular}
\label{tab:comp_cls}
\vspace{-1em}
\end{table}

\begin{table}[tb!]
\centering
\caption{Comparison of source counting accuracy (SCA; \%)}
\vspace{-0.5em}
\scalebox{1}{
\begin{tabular}{c|ccc|c}
\hline
Number of sound sources     & 2             & 3             & 4             & Total           \\ \hline
FUSS loss \cite{wisdom2021s}                   & 0.49          & 0.53          & 0.15          & 0.39          \\
SA-SDR loss \cite{von2022sa}                 & 0.71          & 0.53          & 0.15          & 0.49          \\
CBIR loss \cite{kim2023training}                   & 0.60          & 0.60          & 0.24          & 0.48          \\
\textbf{Cross-Entropy loss} & \textbf{0.77} & \textbf{0.63} & \textbf{0.26} & \textbf{0.55} \\ \hline
\end{tabular}}
\label{tab:comp_sca}
\vspace{-1em}
\end{table}

Finally, the SCA of the proposed classification-based training was compared to conventional threshold-based loss functions. In threshold-based approaches, the estimated source was considered active if the estimated source power was greater than -30 dB relative to the mixture power (Table \ref{tab:comp_sca}). The proposed classification-based source counting method achieved the highest SCA for all number of sources. This suggests that the separation-classification framework has the potential to replace conventional threshold-based methods for source counting. The audio clips and spectrogram examples for the demo are available on the webpage\footnote{\url{https://donghoney0416.github.io/DeFTMamba}}.

\section{Conclusion}
This study presents a unified framework for USS and polyphonic audio classification, improving the performance of both tasks in complex environments. The proposed DeFT-Mamba effectively captures local time-frequency relations and position-wise global relations, surpassing existing SOTA multichannel separation models in tests conducted using the USS dataset designed for simulating in-class polyphony and moving source conditions. By first separating sound sources into individual tracks and then classifying them, the model significantly enhances polyphonic audio classification performance. Moreover, the classification-based source counting method improves source counting accuracy, and separation refinement tuning further boosts overall performance. 

\label{section:references}

\bibliographystyle{IEEEtran}

\begin{thebibliography}{10}
\providecommand{\url}[1]{#1}
\csname url@samestyle\endcsname
\providecommand{\newblock}{\relax}
\providecommand{\bibinfo}[2]{#2}
\providecommand{\BIBentrySTDinterwordspacing}{\spaceskip=0pt\relax}
\providecommand{\BIBentryALTinterwordstretchfactor}{4}
\providecommand{\BIBentryALTinterwordspacing}{\spaceskip=\fontdimen2\font plus
\BIBentryALTinterwordstretchfactor\fontdimen3\font minus \fontdimen4\font\relax}
\providecommand{\BIBforeignlanguage}[2]{{%
\expandafter\ifx\csname l@#1\endcsname\relax
\typeout{** WARNING: IEEEtran.bst: No hyphenation pattern has been}%
\typeout{** loaded for the language `#1'. Using the pattern for}%
\typeout{** the default language instead.}%
\else
\language=\csname l@#1\endcsname
\fi
#2}}
\providecommand{\BIBdecl}{\relax}
\BIBdecl

\bibitem{haykin2005cocktail}
S.~Haykin and Z.~Chen, ``The cocktail party problem,'' \emph{Neural computation}, vol.~17, no.~9, pp. 1875--1902, Sep. 2005.

\bibitem{wang2006computational}
D.~Wang, \emph{Computational auditory scene analysis: principles, algorithms, and applications}.\hskip 1em plus 0.5em minus 0.4em\relax Wiley, 2006.

\bibitem{kavalerov2019universal}
I.~Kavalerov, S.~Wisdom, H.~Erdogan, B.~Patton, K.~Wilson \emph{et~al.}, ``Universal sound separation,'' in \emph{Proc. IEEE Workshop Appl. Signal Process. Audio, Acoust.}, New Paltz, NY, 2019, pp. 175--179.

\bibitem{wisdom2021s}
S.~Wisdom, H.~Erdogan, D.~P. Ellis, R.~Serizel, N.~Turpault, E.~Fonseca \emph{et~al.}, ``What’s all the fuss about free universal sound separation data?'' in \emph{Proc. Int. Conf. Acoust. Speech, Signal Process.}, Tronto, Canada, 2021, pp. 186--190.

\bibitem{tzinis2020improving}
E.~Tzinis, S.~Wisdom, J.~R. Hershey, A.~Jansen, and D.~P. Ellis, ``Improving universal sound separation using sound classification,'' in \emph{Proc. Int. Conf. Acoust. Speech, Signal Process.}, Barcelona, Spain, 2020, pp. 96--100.

\bibitem{liu2024audio}
Y.~Liu, X.~Liu, Y.~Zhao, Y.~Wang, R.~Xia, P.~Tain \emph{et~al.}, ``Audio prompt tuning for universal sound separation,'' in \emph{Proc. Int. Conf. Acoust. Speech, Signal Process.}, Seoul, Korea, 2024, pp. 1446--1450.

\bibitem{pons2024gass}
J.~Pons, X.~Liu, S.~Pascual, and J.~Serr{\`a}, ``Gass: Generalizing audio source separation with large-scale data,'' in \emph{Proc. Int. Conf. Acoust. Speech, Signal Process.}, Seoul, Korea, 2024, pp. 546--550.

\bibitem{luo2023music}
Y.~Luo and J.~Yu, ``Music source separation with band-split {RNN},'' \emph{IEEE/ACM Trans. Audio, Speech, Lang. Process.}, vol.~31, pp. 1893--1901, May. 2023.

\bibitem{fonseca2021fsd50k}
E.~Fonseca, X.~Favory, J.~Pons, F.~Font, and X.~Serra, ``{FSD50K}: an open dataset of human-labeled sound events,'' \emph{IEEE/ACM Trans. Audio, Speech, Lang. Process.}, vol.~30, pp. 829--852, Dec. 2021.

\bibitem{petermann2023tackling}
D.~Petermann, G.~Wichern, A.~S. Subramanian, Z.-Q. Wang, and J.~Le~Roux, ``Tackling the cocktail fork problem for separation and transcription of real-world soundtracks,'' \emph{IEEE/ACM Trans. Audio, Speech, Lang. Process.}, vol.~31, pp. 2592--2605, Jul. 2023.

\bibitem{veluri2023real}
B.~Veluri, J.~Chan, M.~Itani, T.~Chen, T.~Yoshioka, and S.~Gollakota, ``Real-time target sound extraction,'' in \emph{Proc. Int. Conf. Acoust. Speech, Signal Process.}, Rhodes, Greece, 2023, pp. 1--5.

\bibitem{wang2023tf}
Z.-Q. Wang, S.~Cornell, S.~Choi, Y.~Lee, B.-Y. Kim, and S.~Watanabe, ``{TF-GridNet}: Integrating full-and sub-band modeling for speech separation,'' \emph{IEEE/ACM Trans. Audio, Speech, Lang. Process.}, Aug. 2023.

\bibitem{quan2024spatialnet}
C.~Quan and X.~Li, ``{SpatialNet}: Extensively learning spatial information for multichannel joint speech separation, denoising and dereverberation,'' \emph{IEEE/ACM Trans. Audio, Speech, Lang. Process.}, vol.~32, pp. 1310--1323, Feb. 2024.

\bibitem{lee2023deft}
D.~Lee and J.-W. Choi, ``{DeFT-AN}: Dense frequency-time attentive network for multichannel speech enhancement,'' \emph{IEEE Signal Process. Lett.}, vol.~30, pp. 155--159, Feb. 2023.

\bibitem{chen23g_interspeech}
C.~Chen, C.-H.~H. Yang, K.~Li, Y.~Hu, P.-J. Ku, and E.~S. Chng, ``A neural state-space modeling approach to efficient speech separation,'' in \emph{Proc. INTERSPEECH}, Dublin, Ireland, 2023, pp. 3784--3788.

\bibitem{zhao2024sicrn}
C.~Zhao, S.~He, and X.~Zhang, ``{SICRN}: Advancing speech enhancement through state space model and inplace convolution techniques,'' in \emph{Proc. Int. Conf. Acoust. Speech, Signal Process.}, Seoul, Korea, 2024, pp. 10\,506--10\,510.

\bibitem{gu2023mamba}
A.~Gu and T.~Dao, ``Mamba: Linear-time sequence modeling with selective state spaces,'' \emph{arXiv preprint arXiv:2312.00752}, [Online]. Available: \url{https://arxiv.org/abs/2312.00752}, 2023.

\bibitem{li2024spmamba}
K.~Li and G.~Chen, ``{SP}mamba: State-space model is all you need in speech separation,'' \emph{arXiv preprint arXiv:2404.02063}, [Online]. Available: \url{https://arxiv.org/abs/2404.02063}, 2024.

\bibitem{jiang2024dual}
X.~Jiang, C.~Han, and N.~Mesgarani, ``Dual-path mamba: Short and long-term bidirectional selective structured state space models for speech separation,'' \emph{arXiv preprint arXiv:2403.18257}, [Online]. Available: \url{https://arxiv.org/abs/2403.18257}, 2024.

\bibitem{zhang2024mamba}
X.~Zhang, Q.~Zhang, H.~Liu, T.~Xiao, X.~Qian, B.~Ahmed \emph{et~al.}, ``Mamba in speech: Towards an alternative to self-attention,'' \emph{arXiv preprint arXiv:2405.12609}, [Online]. Available: \url{https://arxiv.org/abs/2405.12609}, 2024.

\bibitem{politis2023starss23}
A.~Politis, K.~Shimada, P.~Sudarsanam, A.~Hakala, S.~Takahashi, D.~Krause \emph{et~al.}, ``{STARSS23}: Sony-tau realistic spatial soundscapes 2023,'' \emph{arXiv preprint arXiv:2306.09126}, [Online]. Available: \url{https://arxiv.org/abs/2306.09126}, 2023.

\bibitem{cakir2017convolutional}
E.~Cak{\i}r, G.~Parascandolo, T.~Heittola, H.~Huttunen, and T.~Virtanen, ``Convolutional recurrent neural networks for polyphonic sound event detection,'' \emph{IEEE/ACM Trans. Audio, Speech, Lang. Process.}, vol.~25, no.~6, pp. 1291--1303, May. 2017.

\bibitem{hershey2017cnn}
S.~Hershey, S.~Chaudhuri, D.~P. Ellis, J.~F. Gemmeke, A.~Jansen, R.~C. Moore \emph{et~al.}, ``{CNN} architectures for large-scale audio classification,'' in \emph{Proc. Int. Conf. Acoust. Speech, Signal Process.}, New Orleans, LA, 2017, pp. 131--135.

\bibitem{gong2021ast}
Y.~Gong, Y.-A. Chung, and J.~Glass, ``{AST}: Audio spectrogram transformer,'' in \emph{Proc. INTERSPEECH}, Brno, Czechia, 2021, pp. 571--575.

\bibitem{erol2024audio}
M.~H. Erol, A.~Senocak, J.~Feng, and J.~S. Chung, ``Audio mamba: Bidirectional state space model for audio representation learning,'' \emph{arXiv preprint arXiv:2406.03344}, [Online]. Available: \url{https://arxiv.org/abs/2406.03344}, 2024.

\bibitem{mesaros2019sound}
A.~Mesaros, A.~Diment, B.~Elizalde, T.~Heittola, E.~Vincent, B.~Raj \emph{et~al.}, ``Sound event detection in the dcase 2017 challenge,'' \emph{IEEE/ACM Trans. Audio, Speech, Lang. Process.}, vol.~27, no.~6, pp. 992--1006, Mar. 2019.

\bibitem{lee2023deftan}
D.~Lee and J.-W. Choi, ``{DeFTAN-II}: Efficient multichannel speech enhancement with subgroup processing,'' \emph{arXiv preprint arXiv:2308.15777}, [Online]. Available: \url{https://arxiv.org/abs/2308.15777}, 2023.

\bibitem{dao2024transformers}
T.~Dao and A.~Gu, ``Transformers are {SSMs}: Generalized models and efficient algorithms through structured state space duality,'' in \emph{Proc. Int. Conf. Mach. Learn.}, Vienna, Austria, 2024.

\bibitem{lee24g_interspeech}
D.~Lee and J.-W. Choi, ``{DeFTAN-AA}: Array geometry agnostic multichannel speech enhancement,'' in \emph{Proc. INTERSPEECH}, Kos, Greece, 2024, pp. 3360--3364.

\bibitem{misra2019mish}
D.~Misra, ``Mish: A self regularized non-monotonic activation function,'' in \emph{Proc. Brit. Mach. Vis. Conf.}, Virtual, 2020.

\bibitem{dao2023flashattention}
T.~Dao, ``Flash{A}ttention-2: Faster attention with better parallelism and work partitioning,'' Vienna, Austria, 2024.

\bibitem{chen2020dual}
J.~Chen, Q.~Mao, and D.~Liu, ``Dual-path transformer network: Direct context-aware modeling for end-to-end monaural speech separation,'' in \emph{Proc. INTERSPEECH}, Shanghai, China, 2020, pp. 2642--2646.

\bibitem{wang2021tstnn}
K.~Wang, B.~He, and W.-P. Zhu, ``{TSTNN}: Two-stage transformer based neural network for speech enhancement in the time domain,'' in \emph{Proc. Int. Conf. Acoust. Speech, Signal Process.}, Toronto, Canada, 2021, pp. 7098--7102.

\bibitem{panayotov2015librispeech}
V.~Panayotov, G.~Chen, D.~Povey, and S.~Khudanpur, ``Librispeech: an asr corpus based on public domain audio books,'' in \emph{Proc. Int. Conf. Acoust. Speech, Signal Process.}, Brisbane, Australia, 2015, pp. 5206--5210.

\bibitem{rafii2017musdb18}
Z.~Rafii, A.~Liutkus, F.-R. St{\"o}ter, S.~I. Mimilakis, and R.~Bittner, ``The {MUSDB18} corpus for music separation,'' 10.5281/zenodo.1117371. hal-02190845, Dec. 2017.

\bibitem{gong2022vocalsound}
Y.~Gong, J.~Yu, and J.~Glass, ``Vocalsound: A dataset for improving human vocal sounds recognition,'' in \emph{Proc. Int. Conf. Acoust. Speech, Signal Process.}, Singapore, 2022, pp. 151--155.

\bibitem{diaz2021gpurir}
D.~Diaz-Guerra, A.~Miguel, and J.~R. Beltran, ``gpu{RIR}: A python library for room impulse response simulation with gpu acceleration,'' \emph{Multimedia Tools and Applications}, vol.~80, no.~4, pp. 5653--5671, 2021.

\bibitem{von2022sa}
T.~von Neumann, K.~Kinoshita, C.~Boeddeker, M.~Delcroix, and R.~Haeb-Umbach, ``{SA-SDR}: A novel loss function for separation of meeting style data,'' in \emph{Proc. Int. Conf. Acoust. Speech, Signal Process.}, Singapore, 2022, pp. 6022--6026.

\bibitem{yu2017permutation}
D.~Yu, M.~Kolb{\ae}k, Z.-H. Tan, and J.~Jensen, ``Permutation invariant training of deep models for speaker-independent multi-talker speech separation,'' in \emph{Proc. Int. Conf. Acoust. Speech, Signal Process.}, New Orleans, LA, 2017, pp. 241--245.

\bibitem{kim2023training}
H.~Kim and J.~W. Shin, ``On training speech separation models with various numbers of speakers,'' \emph{IEEE Signal Process. Lett.}, vol.~30, pp. 1202--1206, Aug. 2023.

\bibitem{thiede2000peaq}
T.~Thiede, W.~C. Treurniet, R.~Bitto, C.~Schmidmer, T.~Sporer, J.~G. Beerends \emph{et~al.}, ``{PEAQ}-the {ITU} standard for objective measurement of perceived audio quality,'' \emph{J. Audio Eng. Soc.}, vol.~48, no. 1/2, pp. 3--29, 2000.

\bibitem{gu2024rezero}
R.~Gu and Y.~Luo, ``Re{Z}ero: Region-customizable sound extraction,'' \emph{IEEE/ACM Trans. Audio, Speech, Lang. Process.}, vol.~32, pp. 2576--2589, Apr. 2024.

\end{thebibliography}

\end{document}